\documentclass[peerreview]{IEEEtran}

\usepackage{amssymb}
\usepackage{amsmath}
\usepackage{graphicx}

\newtheorem{thm}{Theorem}
\newtheorem{lemma}[thm]{Lemma}
\newtheorem{definition}[thm]{Definition}
\newtheorem{proposition}[thm]{Proposition}
\newtheorem{corollary}[thm]{Corollary}

\begin{document}

\title {Guessing Based On Length Functions}

\author {Rajesh~Sundaresan,~\IEEEmembership{Senior Member,~IEEE}
\thanks{R. Sundaresan is with the Department of Electrical Communication
Engineering, Indian Institute of Science, Bangalore 560012, India}
\thanks{This work was supported by the Defence Research and Development Organisation,
Ministry of Defence, Government of India, under the DRDO-IISc
Programme on Advanced Research in Mathematical Engineering, and by
the University Grants Commission under Grant Part (2B)
UGC-CAS-(Ph.IV).} }

\maketitle

\begin{abstract}
\label{abs:abstract} A guessing wiretapper's performance on a
Shannon cipher system is analyzed for a source with memory. Close
relationships between guessing functions and length functions are
first established. Subsequently, asymptotically optimal encryption
and attack strategies are identified and their performances analyzed
for sources with memory. The performance metrics are exponents of
guessing moments and probability of large deviations. The metrics
are then characterized for unifilar sources. Universal
asymptotically optimal encryption and attack strategies are also
identified for unifilar sources. Guessing in the increasing order of
Lempel-Ziv coding lengths is proposed for finite-state sources, and
shown to be asymptotically optimal. Finally, competitive optimality
properties of guessing in the increasing order of description
lengths and Lempel-Ziv coding lengths are demonstrated.
\end{abstract}


\section{INTRODUCTION}
\label{sec:introduction}

We consider the classical Shannon cipher system
\cite{194910BSTJ_Sha}. Let $X^n = (X_1, \cdots, X_n)$ be a message
where each letter takes values on a finite set $\mathbb{X}$. This
message should be communicated securely from a transmitter to a
receiver, both of which have access to a common secure key $U^k$ of
$k$ purely random bits independent of $X^n$. The transmitter
computes the cryptogram $Y = f_n(X^n,U^k)$ and sends it to the
receiver over a public channel. The cryptogram may be of variable
length. The function $f_n$ is invertible given $U^k$. The receiver,
knowing $Y$ and $U^k$, computes $X^n = f_n^{-1}(Y,U^k)$. The
functions $f_n$ and $f_n^{-1}$ are published. An attacker
(wiretapper) has access to the cryptogram $Y$, knows $f_n$ and
$f_n^{-1}$, and attempts to identify $X^n$ without knowledge of
$U^k$. The attacker can use knowledge of the statistics of $X^n$. We
assume that the attacker has a test mechanism that tells him whether
a guess $\hat{X}^n$ is correct or not. For example, the attacker may
wish to attack an encrypted password or personal information to gain
access to, say, a computer account, or a bank account via internet,
or a classified database \cite{199909TIT_MerAri}. In these
situations, successful entry into the system or a failure provides
the natural test mechanism. We assume that the attacker is allowed
an unlimited number of guesses. Given the probability mass function
(PMF) of $X^n$, the function $f_n$, and the cryptogram $Y$, the
attacker can determine the posterior probabilities of the message
$P_{X^n|Y}(\cdot \mid y)$. His best guessing strategy having
observed $Y=y$ is then to guess in the decreasing order of these
posterior probabilities $P_{X^n|Y}(\cdot \mid y)$. The key rate for
the system is $k / n = R$ which represents the number of bits of key
used to communicate one message letter.

Merhav and Arikan \cite{199909TIT_MerAri} study discrete memoryless
sources (DMS) in the above setting and characterize the best
attainable moments of the number of guesses that the attacker has to
submit before success. In particular, they show that for a DMS with
the governing single letter PMF $P$ on $\mathbb{X}$, the value of
the optimal guessing exponent is given by
\[
  E(R, \rho) = \max_{Q} \left[ \rho \min \{ H(Q), R \} - D(Q \parallel P)
  \right],
\]
where the maximization is over all PMFs $Q$ on $\mathbb{X}$, $H(Q)$
is the Shannon entropy of the PMF $Q$, and $D(Q \parallel P)$ is the
Kullback-Leibler divergence between $Q$ and $P$. They also show that
$E(R, \rho)$ equals $\rho R$ for $R < H(P)$, and equals the constant
$\rho H_{1/(1+\rho)}(P)$ for $R > H(P_{\rho})$. When $R < H(P)$, the
key rate is not sufficiently large, and an exhaustive key-search
attack is asymptotically optimal. When $R > H(P_{\rho})$, the
randomness introduced by the key is near perfect, and the cryptogram
is useless to the attacker. The attacker submits guesses based
directly on the message statistics, and $\rho H_{1/(1+\rho)}(P)$ is
known to be the optimal guessing exponent in this scenario
\cite{Arikan}, where $H_{1/(1+\rho)}(P)$ is the R\'{e}nyi entropy of
the DMS $P$. For $H(P) < R < H(P_{\rho})$, the optimal strategy
makes use of both the key and the message statistics. $P_{\rho}$ is
the PMF of an auxiliary DMS given by (\ref{eqn:tilted}). Merhav and
Arikan \cite{199909TIT_MerAri} also determine the best achievable
performance based on the large deviations of the number of guesses
for success, and show that it equals the Fenchel-Legendre transform
of $E(R,\rho)$ as a function of $\rho$.

Secret messages typically come from the natural languages which can
be well-modeled as sources with memory, for e.g., a Markov source of
an appropriate order. In this paper, we extend the results of Merhav
and Arikan \cite{199909TIT_MerAri} to sources with memory. As a
first step towards this, we first consider the perfect secrecy
scenario (for e.g., those analogous to $R \geq H(P_{\rho})$ in the
DMS case), and identify a tight relationship between the number of
guesses for success and a lossless source coding length function.
Specifically, we sandwich the number of guesses on either side by a
suitable length function. Arikan's result \cite{Arikan} that the
best value of the guessing exponent for memoryless sources is the
R\'{e}nyi entropy of an appropriate order immediately follows by
recognizing that it is the least value of an average exponential
coding length problem proposed and solved by Campbell
\cite{Campbell-1}. Our approach based on length functions has the
benefit of showing that guessing in the increasing order of lengths
of compressed strings can yield a good attack strategy for sources
with memory. In particular, guessing in the increasing order of
Lempel-Ziv code lengths \cite{197809TIT_ZivLem} for finite-state
sources and increasing description lengths for unifilar sources
\cite{199105TIT_Mer} are asymptotically optimal in a sense made
precise in the sequel.

Next, we establish similar connections between guessing and source
compression for the key-constrained scenarios (i.e., those analogous
to $R < H(P_{\rho})$ in the memoryless case). We then study guessing
exponents for the cipher system on sources with memory, and then
specialize our results to show that all conclusions of Merhav and
Arikan in \cite{199909TIT_MerAri} for memoryless sources extend to
unifilar sources. We also consider the large deviations performance
of the number of guesses and show that attacks based on the
Lempel-Ziv coding lengths and minimum description lengths are
asymptotically optimal for finite-state and unifilar sources,
respectively. We then establish competitive optimality results for
guessing based on these two length functions.

The paper is organized as follows. In Section
\ref{sec:perfectEncryption} we study guessing under perfect secrecy
and establish the relationship between guessing and source
compression. In Section \ref{sec:specifiedKeyRateGuessing}, we study
the key-rate constrained system, establish optimal strategies for
both parties for sources with memory, and study the relationship
between guessing and a new source coding problem. In Section
\ref{sec:unifilarSources}, we characterize the performance for
unifilar sources. In Section \ref{sec:largeDeviations}, we study the
large deviations performance and establish the optimality properties
of guessing based on Lempel-Ziv and minimum description lengths.
Section \ref{sec:concludingRemarks} summarizes the paper and
presents some open problems.

\section{Guessing under perfect secrecy and source compression}

\label{sec:perfectEncryption}

Let us first consider the following ideal setting where $k = nR \geq
n \log |\mathbb{X}|$. Enumerate all the sequences in $\mathbb{X}^n$
from 0 to $|\mathbb{X}|^n-1$ and let the function $f_n$ be the
bit-wise XOR of the key bits and the bits representing the index of
the message. The cryptogram is the message whose index is the output
of $f_n$. The decryption function is also clear - simply XOR the
bits representing the cryptogram with the key bits. Such an
encryption renders the cryptogram completely useless to an attacker
who does not have knowledge of the key. The attacker's optimal
strategy is to guess the message in the decreasing order of message
probabilities. In case the attacker does not have access to the
message probabilities, a robust strategy is needed. We first relate
the problem of guessing to one of source compression. As we will see
soon, robust source compression strategies lead to robust guessing
strategies.

For ease of exposition, and because we have perfect encryption, let
us assume that the message space is simply $\mathbb{X}$. The
extension to strings of length $n$ is straightforward.

A guessing function
\[
  G : \mathbb{X} \rightarrow \left\{1, 2, \cdots, |\mathbb{X}| \right\}
\]
is a bijection that denotes the order in which the elements of
$\mathbb{X}$ are guessed. If $G(x) = i$, then the $i$th guess is
$x$. A length function
\[
  L : \mathbb{X} \rightarrow \mathbb{N}
\]
is one that satisfies Kraft's inequality
\begin{equation}
  \label{eqn:KraftInequality}
  \sum_{x \in \mathbb{X}} 2^{-L(x)} \leq 1.
\end{equation}
To each guessing function $G$, we associate a PMF $Q_G$ on
$\mathbb{X}$ and a length function $L_G$ as follows.

\begin{definition}
Given a guessing function $G$, we say $Q_G$ defined by
\begin{equation}
  \label{eqn:Q_G}
  Q_G(x) = c^{-1} \cdot G(x)^{-1}, ~\forall x \in \mathbb{X},
\end{equation}
is the PMF on $\mathbb{X}$ associated with $G$. The quantity $c$ in
(\ref{eqn:Q_G}) is the normalization constant. We say $L_G$ defined
by
\begin{equation}
  \label{eqn:L_G}
  L_G(x) = \left\lceil - \log Q_G(x) \right\rceil, ~\forall x \in
  \mathbb{X},
\end{equation}
is the length function associated with $G$.
\hspace*{\fill}~\QEDopen
\end{definition}

Observe that
\begin{equation}
  \label{eqn:c}
  c = \sum_{a \in \mathbb{X}} G(a)^{-1} = \sum_{i=1}^{|\mathbb{X}|} \frac{1}{i} \leq 1 + \ln
  |\mathbb{X}|,
\end{equation}
and therefore the PMF in (\ref{eqn:Q_G}) is well-defined. We record
the intimate relationship between these associated quantities in the
following result.

\begin{proposition}
\label{prop:guessingBounds} Given a guessing function $G$, the
associated quantities satisfy
\begin{eqnarray}
  \label{eqn:guessing-PMFBounds}
  c^{-1} \cdot Q_G(x)^{-1} =
  G(x) \leq Q_G(x)^{-1}, \\
  \label{eqn:guessing-lengthBounds}
  L_G(x) - 1 - \log c \leq \log G(x) \leq L_G(x).
\end{eqnarray}
\hspace*{\fill}~\QEDopen
\end{proposition}
\begin{proof}
The first equality in (\ref{eqn:guessing-PMFBounds}) follows from
the definition in (\ref{eqn:Q_G}), and the second inequality from
the fact that $c \geq 1$.

The upper bound in (\ref{eqn:guessing-lengthBounds}) follows from
the upper bound in (\ref{eqn:guessing-PMFBounds}) and from
(\ref{eqn:L_G}). The lower bound in
(\ref{eqn:guessing-lengthBounds}) follows from
\begin{eqnarray*}
  \log G(x) & = & \log \left( c^{-1} \cdot Q_G(x)^{-1} \right) \\
  & = & - \log Q_G(x) - \log c \\
  & \geq & \left( \lceil - \log Q_G(x) \rceil - 1 \right) - \log c
  \\
  & = & L_G(x) - 1 - \log c.
\end{eqnarray*}
\end{proof}

We now associate a guessing function $G_L$ to each length function
$L$.

\begin{definition} \label{defn:G_L}
Given a length function $L$, we define the associated guessing
function $G_L$ to be the one that guesses in the increasing order of
$L$-lengths. Messages with the same $L$-length are ordered using an
arbitrary fixed rule, say the lexicographic order on $\mathbb{X}$.
We also define the associated PMF $Q_L$ on $\mathbb{X}$ to be
\begin{equation}
  \label{eqn:Q_L}
  Q_L(x) = \frac{2^{-L(x)}}{\sum_{a \in \mathbb{X}}
  2^{-L(a)}}.
\end{equation}
\hspace*{\fill}~\QEDopen
\end{definition}

\begin{proposition}
\label{prop:lengthBounds} For a length function $L$, the associated
PMF and the guessing function satisfy the following:
\begin{enumerate}
\item $G_L$ guesses messages in the decreasing order of $Q_L$-probabilities;
\item
\begin{equation}
  \label{eqn:G_LBounds}
  \log G_L(x) \leq \log Q_L(x)^{-1} \leq L(x).
\end{equation}
\end{enumerate}
\hspace*{\fill}~\QEDopen
\end{proposition}

\begin{proof}
The first statement is clear from the definition of $G_L$ and from
(\ref{eqn:Q_L}).

Letting $1 \{ E \}$ denote the indicator function of an event $E$,
we have as a consequence of statement 1) that
\begin{eqnarray}
  G_L(x) \nonumber
  & \leq & \sum_{a \in \mathbb{X}} 1 \left\{ Q_L(a) \geq Q_L(x) \right\} \nonumber \\
  & \leq & \sum_{a \in \mathbb{X}} \frac{Q_L(a)}{Q_L(x)} \nonumber \\
  \label{eqn:guessingUpperBound}
  & = & Q_L(x)^{-1},
\end{eqnarray}
which proves the left inequality in (\ref{eqn:G_LBounds}). This
inequality was known to Wyner \cite{197203IC_Wyn}.

The last inequality in (\ref{eqn:G_LBounds}) follows from
(\ref{eqn:Q_L}) and Kraft's inequality (\ref{eqn:KraftInequality})
as follows:
\[
  Q_L(x)^{-1} = 2^{L(x)} \cdot \sum_{a \in
  \mathbb{X}} 2^{-L(a)} \leq 2^{L(x)}.
\]
\end{proof}

Let $\{L(x) \geq B \}$ denote the set $\{ x \in \mathbb{X} \mid L(x)
\geq B \}$. We then have the following easy to verify corollary to
Propositions \ref{prop:guessingBounds} and \ref{prop:lengthBounds}.

\begin{corollary}
\label{cor:inclusions} For a given $G$, its associated length
function $L_G$, and any $B \geq 1$, we have
\begin{eqnarray}
  \lefteqn{ \left\{ L_G(x)  \geq B + 1 + \log c \right\} } \nonumber \\
  & & \subseteq \left\{ G(x) \geq 2^B \right\} \nonumber \\
  \label{eqn:containments}
  & & \subseteq \left\{ L_G(x) \geq B \right\}.
\end{eqnarray}

Analogously, for a given $L$, its associated guessing function
$G_L$, and any $B \geq 1$, we have
\begin{equation}
  \label{eqn:GsubsetL}
  \{ G_L(x) \geq 2^B \} \subseteq \{ L(x) \geq B\}.
\end{equation}
\hspace*{\fill}~\QEDopen
\end{corollary}

The inequalities between the associates in
(\ref{eqn:guessing-lengthBounds}) and (\ref{eqn:G_LBounds}) indicate
the direct relationship between guessing moments and Campbell's
coding problem \cite{Campbell-1}, and that the R\'{e}nyi entropies
are the optimal growth exponents for guessing moments. See
(\ref{eqn:G*L*Relation}) below. They also establish a simple and new
result: the minimum expected value of the logarithm of the number of
guesses is close to the Shannon entropy.

We now demonstrate other relationships between guessing moments and
average exponential coding lengths which will be useful in
establishing universality properties.

\begin{proposition}
Let $L$ be any length function on $\mathbb{X}$, $G_{L}$ the guessing
function associated with $L$, $P$ a PMF on $\mathbb{X}$, $\rho \in
(0, \infty)$, $L^*$ the length function that minimizes $\mathbb{E}
\left[ 2^{\rho L^*(X)} \right]$, where the expectation is with
respect to $P$, $G^*$ the guessing function that proceeds in the
decreasing order of $P$-probabilities and therefore the one that
minimizes $\mathbb{E} \left[ G^*(X)^{\rho} \right]$, and $c$ as in
(\ref{eqn:c}). Then
\begin{equation}
  \label{eqn:ratioUpperBound}
  \frac{\mathbb{E} \left[ G_{L}(X)^{\rho} \right] }{\mathbb{E} \left[ G^*(X)^{\rho} \right]}
  \leq \frac{\mathbb{E} \left[ 2^{\rho L(X)} \right] }{\mathbb{E} \left[ 2^{\rho L^*(X)}
  \right]} \cdot 2^{\rho ( 1 + \log c )}.
\end{equation}
Analogously, let $G$ be any guessing function, and $L_G$ its
associated length function. Then
\begin{equation}
  \label{eqn:ratioLowerBound}
  \frac{\mathbb{E} \left[ G(X)^{\rho} \right] }{\mathbb{E} \left[ G^*(X)^{\rho} \right]}
  \geq \frac{\mathbb{E} \left[ 2^{\rho L_G(X)} \right] }{\mathbb{E} \left[ 2^{\rho L^*(X)}
  \right]} \cdot 2^{- \rho ( 1 + \log c )}.
\end{equation}
Also,
\begin{equation}
  \label{eqn:G*L*Relation}
  \left| \frac{1}{\rho} \log \mathbb{E} \left[ G^*(X)^{\rho} \right]
  - \frac{1}{\rho} \log \mathbb{E} \left[ 2^{\rho L^*(X)}
  \right] \right| \leq 1 + \log c.
\end{equation}
\hspace*{\fill}~\QEDopen
\end{proposition}

\begin{proof}
Observe that
\begin{eqnarray}
  \lefteqn{ \mathbb{E} \left[ 2^{\rho L(X)} \right] }
  \nonumber \\
  \label{eqn:2a}
  & \geq & \mathbb{E} \left[ G_{L}(X)^{\rho}
  \right] \\
  & \geq & \mathbb{E} \left[ G^*(X)^{\rho} \right] \nonumber \\
  \label{eqn:2b}
  & \geq & \mathbb{E} \left[ 2^{\rho L_{G^*}(X)} \right] 2^{-\rho ( 1 + \log c )} \\
  \label{eqn:2c}
  & \geq & \mathbb{E} \left[ 2^{ \rho L^*(X)} \right] 2^{ -\rho ( 1 + \log c )},
\end{eqnarray}
where (\ref{eqn:2a}) follows from (\ref{eqn:G_LBounds}), and
(\ref{eqn:2b}) from the left inequality in
(\ref{eqn:guessing-lengthBounds}). The result in
(\ref{eqn:ratioUpperBound}) immediately follows. A similar argument
shows (\ref{eqn:ratioLowerBound}). Finally, (\ref{eqn:G*L*Relation})
follows from the inequalities leading to (\ref{eqn:2c}) by setting
$L = L^*$.
\end{proof}

Thus if we have a length function whose performance is close to
optimal, then its associated guessing function is close to guessing
optimal. The converse is true as well. Moreover, the optimal
guessing exponent is within $1+ \log c$ of the optimal coding
exponent for the length function.

Let us now consider strings of length $n$. Let $\mathbb{X}^n$ denote
the set of messages and consider $n \rightarrow \infty$. It is now
easy to see that universality in the average exponential coding rate
sense implies existence of a universal guessing strategy that
achieves the optimal exponent for guessing. For each source in the
class, let $P_n$ be its restriction to strings of length $n$ and let
$L_n^*$ denote an optimal length function that attains the minimum
value $\mathbb{E} \left[ 2^{\rho L_n^*(X^n)}\right]$ among all
length functions, the expectation being with respect to $P_n$. On
the other hand, let $L_n$ be a sequence of length functions for the
class of sources that does not depend on the actual source within
the class. Suppose further that the length sequence $L_n$ is
asymptotically optimal, i.e.,
\begin{eqnarray*}
  \lefteqn{ \lim_{n \rightarrow \infty} \frac{1}{n \rho} \log \mathbb{E} \left[ 2^{\rho L_n(X^n)} \right] } \\
  & = & \lim_{n \rightarrow \infty} \frac{1}{n \rho} \log \mathbb{E} \left[ 2^{\rho L_n^*(X^n)}
  \right],
\end{eqnarray*}
for every source belonging to the class. $L_n$ is thus ``univeral''
for (i.e., asymptotically optimal for all sources in) the class. An
application of (\ref{eqn:ratioUpperBound}) by denoting $c$ in
(\ref{eqn:ratioUpperBound}) as $c_n$ followed by the observation $(1
+ \log c_n)/n \rightarrow 0$ shows that the sequence of guessing
strategies $G_{L_n}$ is asymptotically optimal for the class, i.e.,
\begin{eqnarray}
  \lefteqn{ \lim_{n \rightarrow \infty} \frac{1}{n \rho} \log \mathbb{E} \left[ G_{L_n}(X^n)^{\rho} \right] } \nonumber \\
  \label{eqn:growthExponent}
  & = & \lim_{n \rightarrow \infty} \frac{1}{n \rho} \log \mathbb{E} \left[ G^*(X^n)^{\rho} \right] \nonumber.
\end{eqnarray}

Arikan and Merhav \cite{Arikan-Merhav} provide a universal guessing
strategy for the class of discrete memoryless sources (DMS). For the
class of unifilar sources with a known number of states, the minimum
description length encoding is asymptotically optimal for Campbell's
coding length problem (see Merhav \cite{199105TIT_Mer}). It follows
as a consequence of the above argument that guessing in the
increasing order of description lengths is asymptotically optimal.
(See also the development in Section \ref{sec:unifilarSources}). The
left side of (\ref{eqn:ratioUpperBound}) is the extra factor in the
expected number of guesses (relative to the optimal value) due to
lack of knowledge of the specific source in class. Our prior work
\cite{200701TIT_Sun} characterizes this loss as a function of the
uncertainty class.

\section{Guessing with key-rate constraints and source compression}

\label{sec:specifiedKeyRateGuessing}

We continue to consider strings of length $n$. Let $X^n$ be a
message and $U^k$ the secure key of purely random bits independent
of $X^n$. Recall that the transmitter computes the cryptogram $Y =
f_n(X^n,U^k)$ and sends it to the receiver over a public channel.
Given a PMF of $X^n$, the function $f_n$, and the cryptogram $Y$,
the attacker's optimal strategy is to guess in the decreasing order
of posterior probabilities $P_{X^n|Y}(\cdot \mid y)$. Let us denote
this optimal attack strategy as $G_{f_n}$. The key rate for the
system is $k / n = R < \log |\mathbb{X}|$. If the attacker does not
know the source statistics, a robust guessing strategy is needed.
The following is a first step in this direction.

\begin{proposition}
\label{prop:key:LengthBasedGuessing} Let $L_n$ be an arbitrary
length function on $\mathbb{X}^n$. There is a guessing list $G$ such
that for any encryption function $f_n$, we have
\[
  G(x^n \mid y) \leq 2 \min \left\{ 2^{nR}, 2^{L_n(x^n)} \right\}.
\]
\hspace*{\fill}~\QEDopen
\end{proposition}
\begin{proof}
We use a technique of Merhav and Arikan \cite{199909TIT_MerAri}. Let
$G_{L_n}$ denote the associated guessing function that proceeds in
the increasing order of the lengths and completely ignores the
cryptogram. Let $G_{L_n}$ proceed in the order $x_1^n, x_2^n,
\cdots$. By Proposition \ref{prop:guessingBounds}, we need at most
$2^{L_n(x^n)}$ guesses to identify $x^n$.

Consider the alternative exhaustive key-search attack defined by the
following guessing list:
\[
  f_n^{-1}\left( y,u_1^k \right), f_n^{-1}\left( y,u_2^k \right), \cdots,
\]
where $u_1^k, u_2^k, \cdots$ is an arbitrary ordering of the keys.
This strategy identifies $x^n$ in at most $2^{nR}$ guesses.

Finally, let $G(\cdot \mid y)$ be the list that alternates between
the two lists, skipping those already guessed, i.e., the one that
proceeds in the order
\begin{equation}
  \label{eqn:bestMixedStrategy}
  \left\{ x_1^n, f_n^{-1}\left( y,u_1^k \right), x_2^n, f_n^{-1}\left( y,u_2^k \right), \cdots \right\}.
\end{equation}
Clearly, for every $x^n$, we need at most twice the minimum of the
two original lists.
\end{proof}

We now look at a weak converse to the above in the expected sense.
Our proof also suggests an asymptotically optimal encryption
strategy for sources with memory.
\begin{proposition}
\label{prop:key:optimalEncryption} Fix $n \in \mathbb{N}$, $\rho >
0$, and let $c_n$ denote the constant in (\ref{eqn:c}) as a function
of $n$ with $\mathbb{X}^n$ replacing $\mathbb{X}$. There is an
encryption function $f_n$ and a length function $L_n$ such that
every guessing strategy $G(\cdot \mid y)$ (and in particular
$G_{f_n}$) satisfies
\begin{eqnarray*}
  \lefteqn{ \mathbb{E} \left[ G(X^n \mid Y)^{\rho} \right] } \\
  & \geq & \frac{1}{(2 c_n)^{\rho} (2 +
  \rho)} \mathbb{E} \left[ \left( \min \left\{ 2^{L_n\left(X^n\right)}, 2^{nR}
  \right\} \right)^{\rho}
  \right].
\end{eqnarray*}
\hspace*{\fill}~\QEDopen
\end{proposition}
\begin{proof}
The proof is an extension of Merhav and Arikan's proof of \cite[Th.
1]{199909TIT_MerAri} to sources with memory. The idea is to identify
an encryption mechanism that maps messages of roughly equal
probability to each other.

Let $P_n$ be any PMF on $\mathbb{X}^n$. Enumerate the elements of
$\mathbb{X}^n$ in the decreasing order of their probabilities. For
convenience, let $M = 2^{nR}$. If $M$ does not divide
$|\mathbb{X}|^n$, append a few dummy messages of zero probability to
make the number of messages $N$ a multiple of $M$. Index the
messages from 0 to $N-1$. Henceforth, we identify a message by its
index.

Divide the messages into groups of $M$ so that message $m$ belongs
to group $T_j$, where $j = \lfloor m/M \rfloor$, and $\lfloor \cdot
\rfloor$ is the floor function. Enumerate the key streams from 0 to
$M-1$, so that $0 \leq u \leq M-1$. The function $f_n$ is now
defined as follows. For $m = jM + i$ set
\[
  f_n(jM+i, u) \stackrel{\Delta}{=} jM + \left( i \oplus u \right),
\]
where $i \oplus u$ is the bit-wise XOR operation. Thus messages in
group $T_j$ are encrypted to messages in the same group. The index
$i$ identifying the specific message in group $T_j$, i.e., the last
$nR$ bits of $m$, are encrypted via bit-wise XOR with the key
stream. Given $u$ and the cryptogram, decryption is clear -- perform
bit-wise XOR with $u$ on the last $nR$ bits of $y$.

Given a cryptogram $y$, the only information that the attacker
gleans is that the message belongs to the group determined by $y$.
Indeed, if $y \in T_j$
\[
  P_n \left\{ Y = y \right\} = \frac{1}{M} P_n \left\{ X^n \in T_j \right \}
\]
and therefore
\[
  P_n \left\{ X^n = m \mid Y=y \right\} = \left\{
    \begin{array}{ll}
     \frac{ P_n \left\{ X^n = m \right \} }{ P_n \left\{ X^n \in T_j \right\}
     }, & \lfloor m/M \rfloor = j, \\
     0, & \mbox{otherwise},
    \end{array}
  \right.
\]
decreases with $m$ for $m \in T_j$, and is 0 for $m \notin T_j$. The
attacker's best strategy $G_{f_n}(\cdot \mid y)$ is therefore to
restrict his guesses to $T_j$ and guess in the order $jM, jM+1,
\cdots, jM + M-1$. Thus, when $x^n=jM+i$, the optimal attack
strategy requires $i+1$ guesses.

We now analyze the performance of this attack strategy as follows.
\begin{eqnarray}
  \nonumber
  \lefteqn{ \mathbb{E} \left[ G_{f_n}(X^n|Y)^{\rho} \right] } \\
    \nonumber
    &  =   & \sum_{j=0}^{N/M-1} \sum_{i=0}^{M-1} P_n \{ X^n = j M + i \} (i+1)^{\rho} \\
    \label{eqn:sumProbM}
    & \geq & \sum_{j=0}^{N/M-1} \sum_{i=0}^{M-1} P_n \{ X^n = (j+1)M - 1 \} (i+1)^{\rho} \\
    \label{eqn:sumIntBound}
    & \geq & \sum_{j=0}^{N/M-1} P_n \{ X^n = (j+1)M - 1 \} \frac{ M^{1+ \rho}} {1+\rho} \\
    \nonumber
    & \geq & \frac{1}{1+\rho} \sum_{j=0}^{N/M-1} \sum_{i=0}^{M-1} P_n \{ X^n = (j+1)M + i \} M^{\rho} \\
    \label{eqn:sumProbMBound}
        & & \\
    \label{eqn:G*lowerBound}
    & =    & \frac{1}{1+\rho} \sum_{m=M}^{N-1} P_n \{ X^n = m \} M^{\rho}
\end{eqnarray}
where (\ref{eqn:sumProbM}) follows because the arrangement in the
decreasing order of probabilities implies that
\[
  P_n \{ X^n = jM + i \} \geq P_n \{ X^n = (j+1)M - 1 \}
\]
for $i = 0, \cdots, M-1$. Inequality (\ref{eqn:sumIntBound}) follows
because
\[
  \sum_{i=0}^{M-1} (i+1)^{\rho} = \sum_{i=1}^{M} i^{\rho} \geq
  \int_0^M z^{\rho}~dz = \frac{M^{1+\rho}}{1+\rho},
\]
(\ref{eqn:sumProbMBound}) follows because by the decreasing
probability arrangement
\[
  P_n \{X^n = (j+1)M - 1 \} \geq \frac{1}{M} \sum_{i=0}^{M-1} P_n \{ X^n = (j+1)M + i\}.
\]
Thus (\ref{eqn:G*lowerBound}) implies that
\begin{eqnarray}
  \nonumber
  \lefteqn{ \sum_{m=0}^{N-1} P_n \{ X^n = m \} \left( \min \{ m+1, M \}
  \right)^{\rho} } \\
  \nonumber
  & = & \sum_{m=0}^{M-1} P_n \{ X^n = m \} (m+1)^{\rho} + \sum_{m=M}^{N-1} P_n \{ X = m \}
  M^{\rho} \\
  \nonumber
  & \leq & \mathbb{E} \left[ G_{f_n}(X^n|Y)^{\rho} \right] + (1 + \rho) \mathbb{E} \left[ G_{f_n}(X^n|Y)^{\rho}
  \right] \\
  \label{eqn:expectedG*Bound}
  & = & (2 + \rho) \mathbb{E} \left[ G_{f_n}(X^n|Y)^{\rho} \right],
\end{eqnarray}
Set $G_P$ to be the guessing function that guesses in the decreasing
order of $P$-probabilities without regard to $Y$, i.e., $G_P(m) =
m+1$. Let $L_{G_P}$ be the associated length function. Now use
(\ref{eqn:expectedG*Bound}) and (\ref{eqn:guessing-lengthBounds}) to
get
\begin{eqnarray*}
  \lefteqn{ \mathbb{E} \left[ G_{f_n}(X^n|Y)^{\rho} \right] } \\
  & \geq &
  \frac{1}{2+\rho} \mathbb{E} \left[ \left( \min \left\{ G_P(X^n), M \right\} \right)^{\rho}
  \right] \\
  & \geq &
  \frac{1}{2+\rho} \mathbb{E} \left[ \left( \min \left\{ \frac{2^{L_{G_P}(X^n)}}{2 c_n}, M \right\} \right)^{\rho}
  \right] \\
  & \geq &
  \frac{1}{(2 c_n)^{\rho} (2+\rho)} \mathbb{E} \left[ \left( \min \left\{ 2^{L_{G_P}(X^n)}, M \right\} \right)^{\rho}
  \right].
\end{eqnarray*}
Since $G_{f_n}$ is the strategy that minimizes $\mathbb{E} \left[
G(X^n \mid Y)^{\rho} \right]$ , the proof is complete.
\end{proof}

For a given $\rho > 0$, key rate $R>0$, encryption function $f_n$,
define
\[
  E_n(R, \rho)
  \stackrel{\Delta}{=} \sup_{f_n} \frac{1}{n} \log \mathbb{E} \left[
  G_{f_n}(X^n \mid Y)^{\rho} \right].
\]
Propositions \ref{prop:key:LengthBasedGuessing} and
\ref{prop:key:optimalEncryption} naturally suggest the following
coding problem: identify
\begin{equation}
  \label{eqn:newCodingProblem}
  E_{n,l}(R, \rho) \stackrel{\Delta}{=}  \min_{L_n} \frac{1}{n} \log \mathbb{E} \left[ \left( \min \left\{ 2^{L_n(X^n)},
2^{nR} \right\} \right)^{\rho} \right].
\end{equation}
Analogous to (\ref{eqn:G*L*Relation}), we can relate $E_n(R, \rho)$
and $E_{n,l}(R, \rho)$ for a specified key rate $R$. The following
is a corollary to Propositions \ref{prop:key:LengthBasedGuessing}
and \ref{prop:key:optimalEncryption}.

\begin{corollary}
\label{cor:EnEnl} For a given $R, \rho > 0$, we have
\[
  \left| E_{n,l}(R,\rho) - E_n(R,\rho) \right| \leq \frac{\log (2^{2\rho}
  c_n^{\rho} (2+\rho))}{n}.
\]
\hspace*{\fill}~\QEDopen
\end{corollary}

\begin{proof}
Let $L_n^*$ be the length function that achieves $E_{n,l}(R,\rho)$.
By Proposition \ref{prop:key:LengthBasedGuessing}, and after taking
expectations, we have the guessing strategy $G(\cdot \mid y)$ that
satisfies
\begin{eqnarray*}
  \lefteqn{  \mathbb{E} \left[ \left( \min \left\{ 2^{L_n^*(X^n)},
  2^{nR} \right\} \right)^{\rho} \right] } \\
  & \geq & \sup_{f_n} \frac{1}{2^{\rho}} \mathbb{E} \left[ G(X^n \mid Y)^{\rho}
  \right] \\
  & \geq & \sup_{f_n} \frac{1}{2^{\rho}} \mathbb{E} \left[ G_{f_n}(X^n \mid Y)^{\rho}
  \right] \\
  & \geq & \frac{1}{2^{2\rho}c_n^{\rho} (2 + \rho)} \mathbb{E} \left[ \left( \min \left\{ 2^{L_n(X^n)},
  2^{nR} \right\} \right)^{\rho} \right]
\end{eqnarray*}
for a particular $f_n$ and $L_n$ guaranteed by Proposition
\ref{prop:key:optimalEncryption}
\[
  ~~\geq \frac{1}{2^{2\rho} c_n^{\rho} (2 + \rho)} \mathbb{E} \left[ \left( \min \left\{ 2^{L_n^*(X^n)},
  2^{nR} \right\} \right)^{\rho} \right].
\]
Take logarithms and normalize by $n$ to get the bound.
\end{proof}

The magnitude of the difference between $E_n(R, \rho)$ and
$E_{n,l}(R, \rho)$ vanishes as $n \rightarrow \infty$. Thus, the
problem of finding the optimal guessing exponent is the same as that
of finding the optimal exponent for a coding problem. When $R \geq
\log |\mathbb{X}|$, the coding problem in
(\ref{eqn:newCodingProblem}) reduces to the one considered by
Campbell in \cite{Campbell-1}. Proposition
\ref{prop:key:LengthBasedGuessing} shows that the optimal length
function attaining the minimum in (\ref{eqn:newCodingProblem})
yields an asymptotically optimal attack strategy on the cipher
system. Moreover, the encryption strategy in Proposition
\ref{prop:key:optimalEncryption} is asymptotically optimal.

The following Proposition upper bounds the guessing effort needed to
identify the correct message for sources with memory. A sharper
result analogous to the DMS case is shown later for unifilar
sources.
\begin{proposition} For a given $R, \rho > 0$, we have
\begin{equation}
  \label{eqn:EnUpperBound}
  \limsup_{n \rightarrow \infty} E_n(R, \rho) \leq \min \left\{ \rho R, \limsup_{n \rightarrow \infty} E_n(\rho) \right\},
\end{equation}
where
\[
  E_n(\rho) \stackrel{\Delta}{=} \min_{L_n} \frac{1}{n}
  \log \mathbb{E} \left[ 2^{\rho L_n(X^n)} \right].
\]
\hspace*{\fill}~\QEDopen
\end{proposition}
\begin{proof}
By Corollary \ref{cor:EnEnl}, it is sufficient to show that the
sequence $E_{n,l}(R, \rho)$ is upperbounded by the sequence on the
right side of (\ref{eqn:EnUpperBound}). Let $L_n^*$ be the length
function that minimizes $\mathbb{E} \left[ 2^{\rho L_n(X^n)}
\right]$. Observe that $\min \left\{ 2^{\rho n R}, x \right\}$ is a
concave function of $x$ for a fixed $\rho$ and $R$. Jensen's
inequality then yields
\[
  \mathbb{E} \left[ \min \left\{ 2^{\rho n R}, 2^{\rho L_n^*(X^n)}
  \right\}\right] \leq \min \left\{ 2^{\rho n
  R}, \mathbb{E} \left[ 2^{\rho L_n^*(X^n) } \right] \right\}.
\]
Take logarithms, normalize by $n$, and use the definition of
$E_{n,l}(\rho,R)$ to get
\begin{eqnarray*}
  E_{n,l}(R, \rho)
  & \leq & \frac{1}{n} \log \left( \min \left\{ 2^{\rho n R}, \mathbb{E} \left[ 2^{\rho L_n^*(X^n)}
  \right] \right\} \right) \\
  & = & \min \left\{ \rho R, \frac{1}{n}
  \log \mathbb{E} \left[ 2^{\rho L_n^*(X^n)} \right] \right\}.
\end{eqnarray*}
Now take the limsup as $n \rightarrow \infty$ to complete the proof.
\end{proof}

Our results thus far are applicable to a rather general class of
sources with memory. In the next section, we specialize our results
to the important class of unifilar sources. If the source is a DMS
with defining PMF $P$, then the second term within the min in
(\ref{eqn:EnUpperBound}) is known to be $\rho H_{1/(1+\rho)}(P)$,
where $H_{1/(1+\rho)}(P)$ is R\'{e}nyi's entropy of order
$1/(1+\rho)$ for the source. For unifilar sources, we soon show that
the limsup can be replaced by a limit which equals $\rho$ times a
generalization of the R\'{e}nyi entropy for such a source.

\section{Unifilar Sources}

\label{sec:unifilarSources}

In this section, we generalize the DMS results of Merhav and Arikan
\cite{199909TIT_MerAri} to unifilar sources. We first make some
definitions largely following Merhav's notation in
\cite{199105TIT_Mer}.

Let $x^n = (x_1, \cdots, x_n)$ be a string taking values in
$\mathbb{X}^n$. The string $x^n$ needs to be guessed. Let $s^n =
(s_1, \cdots, s_n)$ be another sequence taking values in
$\mathbb{S}^n$ where $|\mathbb{S}| < \infty$. Let $s_0 \in
\mathbb{S}$ be a fixed initial state. A probabilistic source $P_n$
is \emph{finite-state} with $|\mathbb{S}|$ states
\cite{199105TIT_Mer} if the probability of observing the sequence
pair $(x^n, s^n)$ is given by
\[
  P_n(x^n, s^n) = \prod_{i=1}^n P(x_i, s_i \mid s_{i-1}),
\]
where $P(x_i, s_i \mid s_{i-1})$ is the joint probability of letter
$x_i$ and state $s_i$ given the previous state $s_{i-1}$. The
dependence of $P_n$ on the initial state $s_0$ is implicit.
Typically, the letter sequence $x^n$ is observable and the state
sequence $s^n$ is not. Let $H$ denote the entropy-rate of a
finite-state source, i.e.,
\[
  H \stackrel{\Delta}{=} - \lim_{n \rightarrow \infty} n^{-1}
  \sum_{x^n \in \mathbb{X}^n} P_n(x^n) \log P_n(x^n).
\]

A finite-state source is \emph{unifilar} \cite[p.187]{1965xxIT_Ash}
if the state $s_i$ is given by a deterministic mapping
$\phi:\mathbb{X} \times \mathbb{S} \rightarrow \mathbb{S}$ as
\[
  s_i = \phi(x_i, s_{i-1}),
\]
and the mapping $x \mapsto \phi (x, s)$ is one-to-one \footnote{The
definition in \cite{199105TIT_Mer} does not restrict $\phi$ to be
one-to-one.} for each $s \in \mathbb{S}$. Given $s_0$ and the
sequence $x^n$, the state sequence is uniquely determined. Moreover,
given $s_0$ and the state sequence $s^n$, $x^n$ is uniquely
determined. An important example of a unifilar source is a $k$th
order Markov source where $s_i = (x_i, x_{i-1}, \cdots, x_{i-k+1})$.

Fix $x^n \in \mathbb{X}^n$. For $s \in \mathbb{S}, x \in
\mathbb{X}$, let
\[
  Q_{x^n}(x,s) = \frac{1}{n} \sum_{i=1}^n 1\{ x_i = x, s_{i-1} = s \},
\]
where $1\{ A \}$ is the indicator function of the event $A$.
$Q_{x^n}$ is thus an empirical PMF on $\mathbb{S} \times
\mathbb{X}$. Let
\[
  Q_{x^n}(s) = \sum_{x \in \mathbb{X}} Q_{x^n}(x,s).
\]
The use of $Q_{x^n}$ for both the joint and the marginal PMFs is an
abuse of notation. The context should make the meaning clear. Let
\[
  q_{x^n}(x \mid s) = \left\{
   \begin{array}{ll}
     Q_{x^n}(x,s) / Q_{x^n}(s), & Q_{x^n}(s) > 0, \\
     0, & Q_{x^n}(s) = 0
   \end{array}
  \right.
\]
denote the empirical letter probability given the state. (Given that
$\phi$ is one-to-one, this actually defines a transition probability
matrix on the state space). Denote the empirical conditional entropy
as
\[
  H(Q_{x^n}) = - \sum_{s \in \mathbb{S}} \sum_{x \in \mathbb{X}}
  Q_{x^n}(x,s) \log q_{x^n}(x|s),
\]
and the conditional Kullback-Leibler divergence between the
empirical conditional PMF and the one-step transition matrix
$P(x|s)$ as
\[
  D(Q_{x^n} \parallel P) = \sum_{s \in \mathbb{S}} \sum_{x \in \mathbb{X}}
  Q_{x^n}(x,s) \log \frac{q_{x^n}(x \mid s)}{P(x \mid s)}.
\]
Given that we are dealing with multiple random variables, $H(Q)$ and
$D(Q \parallel P)$ usually stand for joint entropy and
Kullback-Leibler divergence of a pair of joint distributions. We
however alert the reader that they stand for conditional values in
our notation.

Let us further define the type $T_{x^n}$ of a sequence $x^n$ as
follows:
\[
  T_{x^n} = \left\{ a^n \in \mathbb{X}^n \mid Q_{a^n} = Q_{x^n}
  \right\}.
\]
For the unifilar source under consideration, it is easy to see that
\begin{equation}
  \label{eqn:typeProbability}
  P_n(x^n) = 2^{-n \left( H(Q_{x^n}) + D ( Q_{x^n} \parallel
  P)\right)},
\end{equation}
i.e., all elements of the same type have the same probability.
Moreover, for a fixed type $Q_{x^n}$, if we set $P(x \mid s) =
q_{x^n}(x \mid s)$ and observe that for the resulting unifilar
source matched to $x^n$, we have $1 \geq P_n \{ T_{x^n} \} =
|T_{x^n}| P_n(x^n)$, we easily deduce from
(\ref{eqn:typeProbability}) that
\begin{equation}
  \label{eqn:typeUpperBound}
  |T_{x^n}| \leq 2^{n H(Q_{x^n})}.
\end{equation}
Consequently, for any unifilar $P_n$,
\begin{equation}
  \label{eqn:typeProbabilityUpperBound}
  P_n \{ T_{x^n} \} \leq 2^{-n D(Q_{x^n} \parallel P)}.
\end{equation}

Using the fact that the mapping $x \mapsto \phi(x,s)$ is one-to-one
for each $s$, it is possible to get the following useful lower
bounds on the size and probability of a type for unifilar sources.
\begin{lemma}(Merhav \cite[Lemma 1]{199105TIT_Mer}, Gutman \cite[Lemma
1]{198903TIT_Gut}) For a unifilar source, there exists a sequence
$\varepsilon(n) = \Theta(n^{-1} \log n)$ such that
\begin{equation}
  \label{eqn:typeLowerBound}
  \left| \frac{1}{n} \log P_n \left\{ T_{x^n} \right\} + D(Q_{x^n} \parallel P) \right| \leq \varepsilon(n)
\end{equation}
for every $x^n \in \mathbb{X}^n$.
\hspace*{\fill}~\QEDopen
\end{lemma}
Consequently, we also have (\cite[eqn. (17)]{199105TIT_Mer}):
\begin{equation}
  \label{eqn:cardTbounds}
  \left| \frac{1}{n} \log \left| T_{x^n} \right| - H(Q_{x^n})
  \right| \leq \varepsilon(n).
\end{equation}

Let us now define in a fashion analogous to the DMS case
\begin{equation}
  \label{eqn:ERrho}
  E(R, \rho) \stackrel{\Delta}{=} \max_Q \left[ \rho h(Q, R) - D(Q \parallel P) \right]
\end{equation}
where $h(Q,R) = \min \{ H(Q), R \}$, $Q$ is a joint PMF on
$\mathbb{S} \times \mathbb{X}$ with letter probabilities given the
state identified by $q(x \mid s)$, and $H(Q)$ is the conditional
entropy
\[
  H(Q) = - \sum_{s \in \mathbb{S}} \sum_{x \in \mathbb{X}} Q(x,s)
  \log q(x \mid s).
\]
$P(x|s)$ is the conditional PMF that defines the unifilar source.
The string $s_0$ is irrelevant in the definition of $E(R,\rho)$.

We now state and prove a generalization of the Merhav and Arikan
result \cite[Th. 1]{199909TIT_MerAri}.

\begin{thm}
\label{thm:unifilar} For any unifilar source, any $\rho > 0$,
\[
  \lim_{n \rightarrow \infty} E_n(R, \rho) = E(R, \rho).
\]
\hspace*{\fill}~\QEDopen
\end{thm}
\begin{proof}
We show that the limiting value of $E_{n,l}(R, \rho)$ exists for the
corresponding coding problem and equals $E(R, \rho)$. Corollary
\ref{cor:EnEnl} then implies that $E_n(R, \rho)$ for the guessing
problem has the same limiting value.

Let $L_n$ be a minimal length function that attains
$E_{n,l}(R,\rho)$. Arrange the elements of $\mathbb{X}^n$ in the
decreasing order of their probabilities. Furthermore, ensure that
all sequences belonging to the same type occur together. Enumerate
the sequences from 0 to $|\mathbb{X}|^n-1$. Henceforth we refer to a
message by its index.

We claim that we may assume $L_n$ is a nondecreasing function of the
message index. Suppose this is not the case. Let $j$ be the first
index where the nondecreasing property is violated, i.e. $L_n(i)
\leq L_n(i+1)$ for $i = 1, \cdots, j-1$, and $L_n(j) > L_n(j+1)$.
Identify the smallest index $j^*$ that satisfies $L_n(j^*) >
L_n(j+1)$. Modify the lengths as follows: set $L'_n(j^*) =
L_n(j+1)$, then $L'_n(i+1) = L_n(i)$ for $i=j^*, \cdots, j$, and
leave the rest unchanged. Call the new set of lengths $L_n$. In
effect, we have ``bubbled'' $L_n(j+1)$ towards the smaller indices
to the nearest location that does not violate the nondecreasing
condition. The new set of lengths will have the same or lower
$\mathbb{E} \left[ \left( \min \{ 2^{L_n(X^n}, 2^{nR} \}
\right)^{\rho} \right]$. By the optimality of the original set of
lengths, the new lengths are also optimal. Furthermore, as a
consequence of the modification, the location of the first index
where $L_n(i) \nleq L_n(i+1)$ has strictly increased. Continue the
process until it terminates; it will after a finite number of steps.
The resulting $L_n$ is nondecreasing and optimal.

Next, observe that
\begin{equation}
  \label{eqn:L_nLowerBound}
  2^{L_n(i)} \geq i + 1
\end{equation}
because the length functions are such that the sequences are
uniquely decipherable. Another way to see (\ref{eqn:L_nLowerBound})
is to observe that index $i$ is the $i+1$st guess when guessing in
the increasing order of $L_n$ as prescribed by the indices, and
therefore (\ref{eqn:G_LBounds}) implies (\ref{eqn:L_nLowerBound}).

We then have the following sequence of inequalities
\begin{eqnarray}
  \nonumber
  \lefteqn{ \sum_{a^n \in \mathbb{X}^n} P_n(a^n) \left( \min \left\{ 2^{L_n(a^n)}, 2^{nR}
  \right\} \right)^{\rho} } \\
  \label{eqn:3a}
  & \geq & P_n(x^n)
  \sum_{a^n \in T_{x^n}} \left( \min \left\{ 2^{L_n(a^n)}, 2^{nR}
  \right\} \right)^{\rho} \\
  \label{eqn:3b}
  & \geq & P_n(x^n) \sum_{i = i_0(T_{x^n})}^{i_0(T_{x^n}) + |T_{x^n}| - 1} \left( \min \left\{ i+1, 2^{nR}
  \right\} \right)^{\rho} \\
  \label{eqn:3c}
  & \geq & P_n(x^n) \sum_{i = 1}^{|T_{x^n}|} \left( \min \left\{ i, 2^{nR}
  \right\} \right)^{\rho} \\
  \nonumber
  & \geq & P_n(x^n) \int_0^{|T_{x^n}|} \left( \min \left\{ y, 2^{nR}
  \right\} \right)^{\rho} ~ dy \\
  \label{eqn:3d}
  & \geq & P_n(x^n) |T_{x^n}| \frac{1}{1+\rho} \left( \min \left\{ |T_{x^n}|, 2^{nR}
  \right\} \right)^{\rho} \\
  \label{eqn:3e}
  & \geq & P \{ T_{x^n} \} \frac{1}{1+\rho} \left( \min \left\{ 2^{n H(Q_{x^n}) - n \varepsilon(n) }, 2^{nR}
  \right\} \right)^{\rho} \\
  \label{eqn:3f}
  & \geq & \frac{2^{-2n \varepsilon(n)}}{1+\rho} 2^{n (\rho \min \{ H(Q_{x^n}), R \} -  D(Q_{x^n} \parallel
  P))},
\end{eqnarray}
where (\ref{eqn:3a}) follows by restricting the sum to sequences in
type $T_{x^n}$, (\ref{eqn:3b}) follows because of
(\ref{eqn:L_nLowerBound}) and by setting $i_0(T_{x^n})$ as the
starting index of type $T_{x^n}$. We can do this because our
ordering clustered all sequences of the same type. Inequality
(\ref{eqn:3c}) holds because every term under the summation is lower
bounded by the corresponding term on the right side. Inequality
(\ref{eqn:3d}) follows because of the following. For simplicity, let
$|T_{x^n}| = N$ and $2^{nR} = M$. When $N \leq M$,
\[
  \frac{1}{N} \int_0^{N} y^{\rho} ~ dy =
  \frac{N^{\rho}}{1+\rho},
\]
and when $N > M$,
\begin{eqnarray*}
  \lefteqn{\frac{1}{N} \int_0^{N} \left( \min \left\{ y, M
  \right\} \right)^{\rho} ~ dy } \\
  & = & \frac{1}{N}
  \int_0^{M} y^{\rho} ~ dy + \frac{1}{N} \int_M^N M^{\rho} ~ dy \\
  & = & \frac{M}{N} \frac{M^{\rho}}{1+\rho} + \left( 1 -
  \frac{M}{N}\right) M^{\rho} \\
  & \geq & \frac{M^{\rho}}{1+\rho}.
\end{eqnarray*}
Inequality (\ref{eqn:3e}) follows from (\ref{eqn:cardTbounds}) and
(\ref{eqn:3f}) follows from (\ref{eqn:typeLowerBound}).

The type $T_{x^n}$ in (\ref{eqn:3f}) is arbitrary. Moreover, $D(Q
\parallel P)$ and $H(Q)$ are continuous functions of $Q$, and the
set of rational empirical functions $\{ Q_{x^n} \}$ become dense in
the class of unifilar sources with $|\mathbb{S}|$ states and
$|\mathbb{X}|$ alphabets, as $n \rightarrow \infty$. From
(\ref{eqn:3f}) and the above facts, we get $\liminf_{n \rightarrow
\infty}E_{n,l}(R, \rho) \geq E(R, \rho)$.

To show the other direction, we define a universal encoding for the
class of unifilar sources on state space $\mathbb{S}$ with alphabet
$\mathbb{X}$. Given a sequence $x^n$, encode each one of the
$|\mathbb{S}|(|\mathbb{X}|-1)$ source parameters $\{ q_{x^n}(x \mid
s) \}$ estimated from $x^n$. Each parameter requires $\log (n+1)$
bits. Then use $nH(Q_{x^n})$ bits to encode the index of $x^n$
within the type $T_{x^n}$. The resulting description length can be
set to
\[
  L_n^*(x^n) = n H(Q_{x^n}) + |\mathbb{S}|(|\mathbb{X}|-1) \log
  (n+1),
\]
where we have ignored constants arising from integral length
constraints. We call this strategy the minimum description length
coding and $L_n^*$ the minimum description lengths.

$L_n^*$ depends on $x^n$ only through its type $T_{x^n}$. Moreover,
there are at most $(n+1)^{|\mathbb{S}|(|\mathbb{X}|-1)}$ types.
Using these facts, (\ref{eqn:typeUpperBound}), and
(\ref{eqn:typeProbabilityUpperBound}), we get
\begin{eqnarray}
  \lefteqn{ \mathbb{E} \left[ \left( \min \left\{ 2^{L_n^*(X^n)}, 2^{nR} \right\}
  \right)^{\rho}\right] } \\
  & \leq & (n+1)^{(1+\rho) |\mathbb{S}|(|\mathbb{X}|-1)} \\
  && ~~~~ \cdot \max_{T_{x^n} \subseteq \mathbb{X}^n}
  P \{ T_{x^n} \} \min \left\{ 2^{n \rho H(Q_{x^n})}, 2^{n \rho R} \right\} \\
  & \leq & (n+1)^{(1+\rho) |\mathbb{S}|(|\mathbb{X}|-1)} 2^{n E(R, \rho)}.
\end{eqnarray}
Take logarithms and normalize by $n$ to get
\[
  \limsup_{n \rightarrow \infty}E_{n,l}(R, \rho) \leq E(R, \rho).
\]
This completes the proof.
\end{proof}

The minimum description length coding works without knowledge of the
true source parameters. Knowledge of the transition function $\phi$
is sufficient. In the context of guessing, the optimal attack
strategy does not depend on knowledge of the source parameters.
Interlacing the exhaustive key-search attack with the attack based
on increasing description lengths is asymptotically optimal.
Incidentally, the encryption strategy of Merhav and Arikan \cite[Th.
1]{199909TIT_MerAri} uses only type information for encoding, and is
applicable to unifilar sources. The same arguments in the proof of
\cite[Th. 1]{199909TIT_MerAri} go to show that their encryption
strategy is asymptotically optimal for unifilar sources.

Let us define the quantity
\begin{equation}
  \label{eqn:Erho}
  E(\rho) \stackrel{\Delta}{=} \max_{Q} \left[ \rho H(Q) - D(Q \parallel
  P) \right].
\end{equation}
Observe that $E(\rho) = E(R, \rho)$ for $R \geq \log |\mathbb{X}|$,
i.e., $E(\rho)$ determines the guessing exponent under perfect
encryption. The following result identifies useful properties of
these functions.
\begin{proposition}
$E(\rho)$ is a convex function of $\rho$. $E(\rho, R)$ is a convex
function of $\rho$ and a concave function of $R$.
\hspace*{\fill}~\QEDopen
\end{proposition}

\begin{proof}
Equation (\ref{eqn:Erho}) is a maximum of affine functions of $\rho$
and is therefore convex in $\rho$. The same is the case for $E(R,
\rho)$. To see the concavity of $E(R, \rho)$ in $R$, write
(\ref{eqn:ERrho}) as done in \cite[Sec. IV]{199909TIT_MerAri} as
\begin{eqnarray}
  \nonumber
  \lefteqn { E(R, \rho) } \\
  \nonumber
  & = & \max_Q \left[ \rho \min_{0 \leq \theta \leq \rho} \left[ \theta H(Q) + (\rho - \theta) R
  \right]
  - D(Q \parallel P) \right] \\
  \nonumber
  & = & \max_Q \min_{0 \leq \theta \leq \rho} \left[ \theta H(Q) + (\rho - \theta) R
  - D(Q \parallel P) \right] \\
  \label{eqn:4a}
  & = & \min_{0 \leq \theta \leq \rho} \max_Q \left[ \theta H(Q) + (\rho - \theta) R
  - D(Q \parallel P) \right] \\
  \label{eqn:4b}
  & = & \min_{0 \leq \theta \leq \rho} \left[ E(\theta) + (\rho - \theta) R
  )\right].
\end{eqnarray}
The maximization and minimization interchange in (\ref{eqn:4a}) is
justified because the term within square brackets, sum of a scaled
conditional entropy and the negative of a conditional divergence, is
indeed concave in $Q$ and affine in $\theta$. Since (\ref{eqn:4b})
is a minimum of affine functions in $R$, it is concave in $R$.
\end{proof}

It is easy to see the following fact for a unifilar source:
\begin{equation}
  \label{eqn:RenyiUnifilar}
  \lim_{n \rightarrow \infty} \frac{1}{n} \log \left( \sum_{x^n \in
  \mathbb{X}^n} P_n(x^n)^{1/(1+\rho)} \right)^{1+\rho} = E(\rho).
\end{equation}
That the left side in (\ref{eqn:RenyiUnifilar}) is at least as large
as the right side follows from the proof in \cite[Appendix
B]{199105TIT_Mer} and the observation that $\rho H(Q) - D(Q
\parallel P)$ is continuous in $Q$ and that the set of rational
empirical PMFs $Q_{x^n}$ is dense in the set of unifilar sources
with state space $\mathbb{S}$ and alphabet $\mathbb{X}$, as $n
\rightarrow \infty$. The other direction is an easy application of
the method of types. The initial state which is implicit in $P_n$
does not affect the value of the limit (as one naturally expects in
this Markov case). In the memoryless case, i.e., when $s_i = x_i$,
and $P(x|s)$ is independent of $s$, this quantity converges to
$E(\rho) = \rho H_{1/(1+\rho)}(P)$ where $H_{1/(1+\rho)}(P)$ is the
R\'{e}nyi entropy of the DMS $P$ on $\mathbb{X}$.

Analogous to a DMS case, we can characterize the behavior of $E(R,
\rho)$ as a function of $R$ for a particular source $P$.

\begin{proposition} For a given $\rho > 0$ and a unifilar source,
let $E'(\rho)$ exist. Then
\[
  E(R, \rho) = \left\{
    \begin{array}{ll}
      \rho R, & R < H, \\
      (\rho - \theta_0) R + E(\theta_0), & H \leq R \leq E'(\rho), \\
      E(\rho), & R > E'(\rho)
    \end{array}
  \right.
\]
where $\theta_0 \in [0,\rho]$ in the second case.
\hspace*{\fill}~\QEDopen
\end{proposition}

\begin{proof}
Indeed, from (\ref{eqn:4b}) it is clear by the continuity of the
term within square brackets that for all values of $R$, $E(R, \rho)
= (\rho - \theta_0) R + E(\theta_0)$ for some $\theta_0 \in [0,
\rho]$, and the second case is directly proved.

Suppose $R < H$. Then we may choose $Q = P$ in (\ref{eqn:ERrho}) to
get $E(R, \rho) \geq \rho R$. However, (\ref{eqn:EnUpperBound})
indicates that $E(R, \rho) \leq \rho R$, which leads us to conclude
that $E(R, \rho) = \rho R$ when $R < H$.

Next observe that $E(R, \rho) \leq E(\rho)$ is direct for all values
of $R$, and in particular for $R > E'(\rho)$. To show the reverse
direction, (\ref{eqn:4b}) yields
\begin{eqnarray*}
  E(R, \rho) & = & \min_{0 \leq \theta \leq \rho} \left[ E(\theta) + (\rho -
  \theta) R \right] \\
  & = & E(\rho) + \min_{0 \leq \theta \leq \rho} (\rho -
  \theta) \left( R - \frac{E(\rho) - E(\theta)}{\rho -
  \theta}\right).
\end{eqnarray*}
The proof will be complete if we can show that the term within
parentheses is nonnegative for $0 \leq \theta \leq \rho$. This holds
because of the following. By the convexity of $E(\theta)$, the
largest value of $(E(\rho) - E(\theta))/(\rho - \theta)$ for the
given range of $\theta$ is $E'(\rho)$ (see for example, Royden
\cite[Lemma 5.5.16]{1988RA_Roy}), and this is upper bounded by $R$.
\end{proof}

For a DMS, Merhav and Arikan \cite{199909TIT_MerAri} show that
$E'(\rho) = H(P_{\rho})$, where $P_{\rho}$ is the PMF given by
\begin{equation}
  \label{eqn:tilted}
  P_{\rho}(x) = \frac{P(x)^{1/(1+\rho)}}{\sum_{a \in \mathbb{X}}
  P(a)^{1/(1+\rho)}}.
\end{equation}
They also show that $\theta_0$ is the unique solution to $R =
H(P_{\theta})$.

\section{Large Deviations Performance}

\label{sec:largeDeviations}

\subsection{General Sources With Memory}

We now study the problem of large deviations in guessing and its
relation to source compression. Our goal is to extend the large
deviations results of Merhav and Arikan \cite{199909TIT_MerAri} to
sources with memory using the tight relationship between guessing
functions and length functions. We begin with the following general
result.
\begin{proposition}
\label{prop:largeDev}
\begin{enumerate}
\item \label{item:largeDev1} When $B > R > 0$, there is an attack strategy that satisfies
\[
  \sup_{f_n} P_n \left\{ G(X^n \mid Y) \geq 2^{nB} \right\} = 0
\]
for all sufficiently large $n$.
\item \label{item:largeDev2} When $B \leq R$, there is an attack strategy that satisfies
\begin{eqnarray*}
  \lefteqn{ \sup_{f_n} P_n \left\{ G(X^n \mid Y) \geq 2^{nB}
  \right\} } \\
  & \leq & \min_{L_n} P_n \left\{ L_n(X^n) \geq n B - 1 \right\}.
\end{eqnarray*}
\item \label{item:largeDev3} When $B < R$, there is an encryption function $f_n$ such
that
\begin{eqnarray*}
  \lefteqn{ P_n \left\{ G_{f_n} (X^n \mid Y) \geq 2^{nB} \right\} }
  \\
  & \geq &
  \frac{1}{3} \cdot \min_{L_n} P_n \left\{ L_n(X^n) \geq n B +1 + \log c_n
  \right\}.
\end{eqnarray*}
\end{enumerate}
\hspace*{\fill}~\QEDopen
\end{proposition}

{\it Remarks}: When $B = R$, the large deviations behavior of
guessing and coding may differ. If we define
\begin{equation}
  \label{eqn:Fn}
  F_n(R,B) \stackrel{\Delta}{=} \inf_{f_n} \left[ - \frac{1}{n} \log P_n \left\{ G_{f_n} (X^n | Y) \geq 2^{nB} \right\}\right]
\end{equation}
and
\begin{equation}
  \label{eqn:Fnl}
  F_{n,l}(B) \stackrel{\Delta}{=} \max_{L_n} \left[ - \frac{1}{n} \log P_n \left\{ L_n(X^n) \geq 2^{nB}
  \right\}\right],
\end{equation}
then $F_n(R,B) = \infty$ for all sufficiently large $n$ if $R < B$.
When $R>B$, $F_n(R,B)$ is bounded between $F_{n,l}(B - 1/n)$ and
$F_{n,l}(B + (1+ \log c_n)/n))$ ignoring vanishing terms.

\begin{proof}
Observe first that for any encryption function, the strategy
(\ref{eqn:bestMixedStrategy}) requires at most $2^{nR+1}$ guesses.
If $B>R$, $2^{nB} > 2^{nR+1}$ for all sufficiently large $n$, and
therefore
\[
  \sup_{f_n} P_n \left\{ G(X^n | Y) \geq 2^{nB} \right\} = 0.
\]

When $B \leq R$, the same strategy with an optimal $L_n$ that
minimizes $P_n \{ L_n(X^n) \geq nB -1 \}$ requires $G(x^n \mid y)
\leq 2 \min \left\{2^{L(x^n)}, 2^{nR} \right\}$ guesses. Hence
\[
  \left\{ G(x^n \mid y) \geq 2^{nB} \right\} \subseteq \left\{ L_n(x^n) \geq nB - 1 \right\}
\]
and therefore
\[
  P_n \{ G(X^n \mid Y) \geq 2^{nB} \} \leq P_n \{ L_n(X^n) \geq nB-1
  \}.
\]
Since this is true for any encryption function $f_n$, the second
statement follows. The attack $G(\cdot \mid y)$ given by
(\ref{eqn:bestMixedStrategy}) interlaces guesses in the increasing
order of the $L_n$ that attains the minimum in $\min_{L_n} P_n
\left\{ L_n(X^n) \geq nB - 1 \right\}$ with the exhaustive
key-search strategy.

Next, let $B < R$ and consider the encryption strategy given in the
proof of Proposition \ref{prop:key:optimalEncryption} with $N = M
\lceil |\mathbb{X}|^n / M \rceil$ (with dummy messages possibly
appended) and $M = 2^{nR}$. Let $G_{P_n}$ denote guessing in the
increasing order of $P_n$-probabilities. Once again we refer to
messages by their indices. For the optimal guessing strategy
$G_{f_n}$, we have
\begin{eqnarray*}
  \lefteqn{ P_n \left\{ G_{f_n}(X^n \mid Y) \geq 2^{nB} \right\} } \\
  & = & \sum_{j=0}^{N/M-1} \sum_{i = 2^{nB}-1}^{M-1}
  P_n \left\{ X^n = jM+i \right\} \\
  & \geq & \sum_{j=0}^{N/M-1} P_n \left\{ X^n = (j+1) M-1 \right\} \left(M - 2^{nB} \right) \\
  & \geq & \sum_{j=0}^{N/M-1} \sum_{i=0}^{M-1} P_n \left\{ X^n = (j+1) M+i \right\} \frac{M - 2^{nB}}{M} \\
  & = & \left(1 - \frac{2^{nB}}{M} \right) \sum_{m=M}^{N-1} P_n \left\{ X^n = m \right\} \\
  & \geq & \frac{1}{2} \sum_{m=M}^{N-1} P_n \left\{ X^n = m \right\},
\end{eqnarray*}
where the last inequality follows because $B < R$. (When $B=R$, the
lower bound is 0 and this technique does not work). Also, rather
trivially,
\[
  P_n \left\{ G_{f_n}(X^n \mid Y) \geq 2^{nB} \right\} \geq \sum_{m=2^{nB}-1}^{M-1}
  P_n \left\{ X^n = m \right\}.
\]
Putting these together, we get
\begin{eqnarray*}
  \sum_{m=2^{nB}-1}^{N-1} P_n \left\{ X^n = m \right\} & = & P_n \left\{ G_{P_n}(X^n) \geq 2^{nB} \right\} \\
  & \leq & 3 P_n \left\{ G_{f_n}(X^n \mid Y) \geq 2^{nB} \right\}.
\end{eqnarray*}
Since $\{ L_{G_{P_n}}(x^n) \geq nB + 1 + \log c_n \} \subseteq \{
G_{P_n}(x^n) \geq 2^{nB} \}$, we get
\begin{eqnarray*}
  \lefteqn{ P_n \{ G_{f_n}(X^n \mid Y) \geq 2^{nB} \} } \\
  & \geq & \frac{1}{3} \cdot P_n \{ L_{G_{P_n}}(X^n) \geq n B + 1 + \log c_n
  \} \\
  & \geq & \frac{1}{3} \cdot \min_{L_n} P_n \{ L_n(X^n) \geq n B + 1 + \log c_n
  \},
\end{eqnarray*}
and this concludes the proof.
\end{proof}

\subsection{Unifilar Sources}

In this subsection, we specialize the result of Proposition
\ref{prop:largeDev} to unifilar sources.

\begin{corollary}
For a unifilar source,
\[
  F(R,B) \stackrel{\Delta}{=} \lim_{n \rightarrow \infty} F_n(R,B) = \left\{
    \begin{array}{ll} \infty, & B > R, \\
                     F(B), & B <
                     R,
    \end{array}
  \right.
\]
where
\[
F(B) \stackrel{\Delta}{=} \min_{Q : H(Q) \geq B} D(Q \parallel P)
\]
is the source coding error exponent for the unifilar source.
\hspace*{\fill}~\QEDopen
\end{corollary}

\begin{proof} This follows straightforwardly from the remarks
immediately following Proposition \ref{prop:largeDev} if we can show
that $\lim_{n \rightarrow \infty}F_{n,l}(B) = F(B)$ and that $F(B)$
is continuous in $(0, \log |\mathbb{X}|)$. This was proved by Merhav
in \cite[Sec. III]{199105TIT_Mer}.
\end{proof}

We remark that the optimal attack strategy does not depend on the
source parameters. Guessing in the increasing order description
lengths, interlaced with the exhaustive key-search attack is an
asymptotically optimal attack. Furthermore, as is the case for
guessing moments, the encryption strategy of Merhav and Arikan
\cite[Th. 2]{199909TIT_MerAri} is easily verified to be an
asymptotically optimal encryption strategy for unifilar sources when
$B < R$.

$E(R, \rho)$ and $F(R,B)$ for unifilar sources are related via the
Fenchel-Legendre transform, i.e.,
\[
  E(R,\rho) = \sup_{B > 0} \left[ \rho B - F(R,B) \right]
\]
and
\[
  F(R,B) = \sup_{\rho > 0} \left[ \rho B - E(R, \rho) \right].
\]
The proof is identical to that of \cite[Th. 3]{199909TIT_MerAri}
where this result is proved for DMSs.

\subsection{Finite-State Sources}

We now consider the larger class of finite state sources. The
Lempel-Ziv coding strategy \cite{197809TIT_ZivLem} asymptotically
achieves the entropy rate of a finite-state source without knowledge
of the source parameters. It is therefore natural to consider its
use in attacking a cipher system that attempts to securely transmit
a message put out by a finite-state source. Our next goal is to show
that guessing in the increasing order of Lempel-Ziv coding lengths
has an interesting universality property.

Let $U_{LZ}: \mathbb{X}^n \rightarrow \mathbb{N}$ be the length
function for the Lempel-Ziv code \cite{197809TIT_ZivLem}. The
following theorem due to Merhav \cite{199105TIT_Mer} indicates that
the Lempel-Ziv algorithm is asymptotically optimal in achieving the
minimum probability of buffer overflow.

\begin{thm}[Merhav \cite{199105TIT_Mer}]
\label{thm:Merhav} For any length function $L_n$, every finite-state
source $P_n$, every $B_n \in (nH, n \log |\mathbb{X}|)$ where $H$ is
the entropy-rate of the source $P_n$, and all sufficiently large
$n$,
\begin{eqnarray}
  \lefteqn{ P_n \{ U_{LZ}(X^n) \geq B_n + n \varepsilon(n) \} } \nonumber \\
  \label{eqn:MerhavInequality}
  & \leq & ( 1 + \delta(n)) \cdot P_n \{ L_n(X^n) \geq B_n \}
\end{eqnarray}
where $\varepsilon(n) = \Theta(1/\sqrt{\log n})$ is a positive
sequence that depends on $|\mathbb{X}|$ and $|\mathbb{S}|$, and
$\delta(n) = n^2 2^{-n \varepsilon(n)}$.
\hspace*{\fill}~\QEDopen
\end{thm}

{\it Remark}: Merhav's result \cite[Th. 1]{199105TIT_Mer} assumes
that $B_n = nB$ for a constant $B \in (H, \log |\mathbb{X}|)$, but
the proof is valid for any sequence $B_n \in (nH, n\log
|\mathbb{X}|)$.

Let $G_{LZ}$ be the short-hand notation for the more cumbersome
$G_{U_{LZ}}$, the guessing function associated with $U_{LZ}$. Let
$c_n$ be as given in (\ref{eqn:c}) with $\mathbb{X}^n$ replacing
$\mathbb{X}$. Furthermore, for the key-constrained cipher system,
let $G_{LZ}(\cdot \mid y)$ denote the attack of guessing in the
order prescribed by $G_{LZ}$ interlaced with the exhaustive
key-search attack. Observe that $G_{LZ} (\cdot \mid y)$ needs
knowledge of $f_n$.

\begin{thm}
\label{thm:largeDevGuessingLZ} For any guessing function $G_n$,
every finite-state source $P_n$, every $B \in (H, \log
|\mathbb{X}|)$ where $H$ is the entropy-rate of the source $P_n$,
and all sufficiently large $n$,
\begin{eqnarray}
  \lefteqn{ P_n \left\{ n^{-1} \log G_{LZ}(X^n) \geq B + \varepsilon(n) + \gamma(n) \right\} } \nonumber \\
  \label{eqn:largeDeviationsGuessingInequality}
  & \leq & (1 + \delta(n)) \cdot P_n \left\{ n^{-1} \log G_n(X^n) \geq B \right\}
\end{eqnarray}
where $\varepsilon(n)$ and $\delta(n)$ are the sequences in
(\ref{eqn:MerhavInequality}), and $\gamma(n) = (1 + \log c_n)/n =
\Theta(n^{-1} \log n)$.

For the key-rate constrained cipher system, let $B < R$. Then for
any encryption function, we have
\begin{eqnarray}
  \lefteqn{ P_n \left\{ n^{-1} \log G_{LZ}(X^n \mid Y) \geq B + 1/n +\varepsilon(n) + \gamma(n) \right\} } \nonumber \\
  &\leq &3(1 + \delta(n)) \cdot \sup_{f_n} P_n \left\{ n^{-1} \log G_{f_n}(X^n \mid Y) \geq B
  \right\} \nonumber \\
  \label{eqn:key:largeDeviationsGuessingInequality}
  &&
\end{eqnarray}
for all sufficiently large $n$.
\hspace*{\fill}~\QEDopen
\end{thm}

{\it Remark}: Thus the Lempel-Ziv coding strategy provides an
asymptotically optimal universal attack strategy for the class of
finite-state sources, in the sense of attaining the limiting value
of (\ref{eqn:Fn}), if the limit exists.

\begin{proof}
Observe that
\begin{eqnarray}
  \lefteqn{ (1+\delta(n)) P_n \left\{ G_n(X^n) \geq 2^{nB} \right\} } \nonumber \\
  \label{eqn:1a}
  & \geq & (1 + \delta(n)) P_n \left\{ L_{G_n}(X^n) \geq n B + 1 + \log c_n \right\} \\
  \label{eqn:1b}
  & \geq & P_n \left\{ U_{LZ}(X^n) \geq n B + 1 + \log c_n + n \varepsilon(n)
  \right\} \\
  \label{eqn:1c}
  & \geq & P_n \left\{ G_{LZ}(X^n) \geq 2^{nB + n\varepsilon(n) + n\gamma(n)} \right\},
\end{eqnarray}
where (\ref{eqn:1a}) follows from the first inclusion in
(\ref{eqn:containments}), and (\ref{eqn:1b}) from
(\ref{eqn:MerhavInequality}). The last inequality (\ref{eqn:1c})
follows from (\ref{eqn:GsubsetL}). This proves the first part.

To show the second part, we use Proposition \ref{prop:largeDev}.3
and Theorem \ref{thm:Merhav} as follows: for all sufficiently large
$n$,
\begin{eqnarray*}
  \lefteqn{ 3(1+\delta(n)) \sup_{f_n} P_n \left\{ G_{f_n}(X^n \mid Y) \geq 2^{nB} \right\} } \nonumber \\
  & \geq & (1 + \delta(n)) P_n \left\{ L_n(X^n) \geq n B + n\gamma(n)
  \right\} \\
  & \geq & P_n \left\{ U_{LZ}(X^n) \geq n B + n\gamma(n) + n\varepsilon(n) \right\} \\
  & \geq & P_n \left\{ G_{LZ}(X^n \mid Y) \geq 2^{nB + 1 + n \gamma(n) + n \varepsilon(n)} \right\}
\end{eqnarray*}
where the last inequality holds for any arbitrary encryption
function with $G_{LZ}(\cdot \mid y)$ being the interlaced attack
strategy.
\end{proof}

Observe that $\varepsilon(n) + \gamma(n) = \Theta(1 / \sqrt{\log
n})$. For unifilar sources, a result analogous to Theorem
\ref{thm:largeDevGuessingLZ} can be shown with $\varepsilon(n) +
\gamma(n) = \Theta(n^{-1} \log n)$. Guessing for this class of
sources proceeds in the order of increasing description lengths.
This conclusion follows from a result analogous to Theorem
\ref{thm:Merhav} on the asymptotic optimality of minimum description
coding (see Merhav \cite[Sec. III]{199105TIT_Mer}).

\subsection{Competitive Optimality}

\label{subsec:competitiveOptimality}

We now demonstrate a competitive optimality property for $G_{LZ}$.
From \cite[eqn. (28)]{199105TIT_Mer} extended to finite-state
sources, we have for any competing code $L_n$
\begin{eqnarray}
  \lefteqn{ P_n \{ U_{LZ}(X^n) > L_n(X^n) + n \varepsilon(n) \} } \nonumber \\
  \label{eqn:MerhavCompetitiveOptimality}
  & \leq & P_n \{ U_{LZ}(X^n) < L_n(X^n) + n \varepsilon(n) \}
\end{eqnarray}
where $\varepsilon(n) = \Theta((\log \log n)/(\log n))$. From
(\ref{eqn:G_LBounds}) and (\ref{eqn:guessing-lengthBounds}), we get
\[
  U_{LZ}(x^n) \geq \log G_{LZ}(x^n)
\]
and
\[
  \log G(x^n) \geq L_G(x^n) - 1 - \log c_n,
\]
respectively. We therefore conclude that
\begin{eqnarray}
  \lefteqn{ \{\log G_{LZ}(x^n) > \log G(x^n) + n (\varepsilon(n) + \gamma(n)) \} } ~~~~~~ \nonumber \\
  & \subseteq & \{ U_{LZ}(x^n) > L_G(x^n) + n \varepsilon(n) \}
  \nonumber
\end{eqnarray}
and that
\begin{eqnarray}
  \lefteqn{ \{ U_{LZ}(x^n) < L_G(x^n) + n \varepsilon(n)
  \}} ~~~~~~ \nonumber \\
  & \subseteq & \{\log G_{LZ}(x^n) < \log G(x^n) + n (\varepsilon(n) + \gamma(n))
  \}. \nonumber
\end{eqnarray}
From these two inclusions and
(\ref{eqn:MerhavCompetitiveOptimality}), we easily deduce the
following result.

\begin{thm} For any finite-state source and any competing guessing
function $G$, we have
\begin{eqnarray*}
  \lefteqn{ P_n \{ \log G_{LZ}(X^n) > \log G(X^n) + n \varepsilon'(n) \} }\\
  & \leq & P_n \{ \log G_{LZ}(X^n) < \log G(X^n) + n \varepsilon'(n) \}
\end{eqnarray*}
where $\varepsilon'(n) = \varepsilon(n) + \gamma(n)$.
\hspace*{\fill}~\QEDopen
\end{thm}

For unifilar sources, the above sequence of arguments for minimum
description length coding and \cite[eqn. (28)]{199105TIT_Mer} imply
that we may take $\varepsilon'(n) = \Theta( n^{-1} \log n )$.

\section{Concluding Remarks}

\label{sec:concludingRemarks}

In this paper, we studied two measures of cryptographic security
based on guessing, for sources with memory. The first one was based
on guessing moments and the second on large deviations performance
of the number of guesses. We identified an asymptotically optimal
encryption strategy that orders the messages in the decreasing order
of their probabilities, enumerates them, and then encrypts as many
least-significant bits as there are key bits. We also identified an
optimal attack strategy based on a length function that attains the
optimal value for a source coding problem. Both these strategies
need knowledge of the message probabilities.

We then specialized our results to the case of unifilar sources,
gave formulas for computing the two measures of performance, and
argued that the optimal encryption strategy as well as the optimal
attack strategy depended on the source parameters only through the
number of states and letters, i.e., the optimal encryption and
attack strategies are universal for this class.

We also showed that an attack strategy based on the Lempel-Ziv
coding lengths is asymptotically optimal for the class of finite
state sources. Finally, we provided competitive optimality results
for guessing in the order of increasing description lengths and
Lempel-Ziv lengths.

We end this paper with a short list of related open problems.
\begin{itemize}
\item Consider a modification to the encryption technique of
Proposition \ref{prop:key:optimalEncryption} where the messages are
enumerated in the increasing order of their Lempel-Ziv lengths
instead of message probabilities. Does this ordering lead to an
asymptotically optimal encryption strategy? Such a strategy would
not depend on the specific knowledge of source parameters.
\item It would be of interest to see if the results on guessing
moments for unifilar sources can be extended to finite-state
sources.
\item The large deviations behavior of guessing when $B = R$ is not
well-understood and might be worth investigating.
\item As mentioned in \cite{199909TIT_MerAri},
one might wish to consider a scenario where only a noisy
version of the cryptogram is available to the attacker. The
degradation in the attacker's performance could be quantified.
\end{itemize}


\bibliography{GuessingBasedOnLengthFunctions.bbl}

\begin{biographynophoto}{Rajesh Sundaresan}
(S'96-M'2000-SM'2006) received his B.Tech. degree in electronics and
communication from the Indian Institute of Technology, Madras, the
M.A. and Ph.D. degrees in electrical engineering from Princeton
University, NJ, in 1996 and 1999, respectively. From 1999 to 2005,
he worked at Qualcomm Inc., Campbell, CA, on the design of
communication algorithms for WCDMA and HSDPA modems. Since 2005 he
is an Assistant Professor in the Electrical Communication
Engineering department at the Indian Institute of Science,
Bangalore. His interests are in the areas of wireless communication
and information theory.
\end{biographynophoto}

\end{document}